\documentclass[prd,aps,showpacs,floats]{revtex4}
\usepackage{amsmath,amssymb}
\usepackage{graphicx}
\usepackage{epsfig}

\begin{document}

\title[Short Title]{Effects of non-standard neutrino interactions 
on MSW-LMA solution to the solar neutrino problem}
\author{M. M. Guzzo}\email{guzzo@ifi.unicamp.br} 
\author{P. C. de Holanda}\email{holanda@ifi.unicamp.br}
\author{O.  L. G. Peres}\email{orlando@ifi.unicamp.br} 
\affiliation{Instituto de F\'\i sica Gleb Wataghin - UNICAMP, 
  13083-970 Campinas SP, Brazil} 
\newcommand{\lsim}{\,\lower .5ex\hbox{$\buildrel < \over {\sim}$}\,}
\newcommand{\gsim}{\,\lower .5ex\hbox{$\buildrel > \over {\sim}$}\,}
\newcommand{\vect}[1]{\overrightarrow{\sf #1}}
\newcommand{\system}[1]{\left\{\matrix{#1}\right.}
\newcommand{\displayfrac}[2]{\frac{\displaystyle #1}{\displaystyle #2}}
\newcommand{\nucl}[2]{{}^{#1}\mbox{#2}}
\newcommand{\diff}{{\rm\,d}}
\newcommand{\ea}{{\em et al.}}
\newcommand{\be}{\begin{equation}}
\newcommand{\ee}{\end{equation}}
\begin{abstract}
We show that the non-standard neutrino interactions can play a role as
sub-leading effect on the solar neutrino oscillations.
We observe that very small flavor universality violations of order of
0.1-0.2 $G_F$ is sufficient to induce two phenomena: suppression of
the $\nu_e$-earth regeneration and a shift of the resonance layer in
the sun. We obtain these phenomena even in the absence of any flavor
changing interactions. We discuss their consequences and confront with
a global analysis of solar+KamLAND results. We conclude that a new
compatibility region in the $\Delta m^2 \times \tan^2 \theta_{\odot}$ , which
we call very low Large Mixing Angle region is found  for 
$\Delta m^2 \sim 10^{-5}$ eV$^2$ and $\tan^2 \theta_{\odot}= 0.45$.

\end{abstract}

\maketitle

\baselineskip=20pt

\section{Introduction}

In the last years, the discovery of neutrino oscillation in
solar and reactor experiments selected as a more probable explanation
to the solar neutrino problem the
so called Large Mixing Angle (LMA) MSW solution. The
SNO~\cite{sno,sno-all} and the KamLAND~\cite{Kam}
experiments confirm and refine the trend of the evidences of neutrino
oscillations due the solar neutrino
observations, as measured by Homestake~\cite{Cl},
SAGE~\cite{sage}, GALLEX~\cite{gallex}, GNO~\cite{gno} and
Super-Kamiokande~\cite{SK,SK2}. As a result, the solar oscillation parameters 
have pinned down to $ 6\times 10^{-5}$ eV$^2 <\Delta m^2 < 1\times 
10^{-4}$ eV$^2$ 
and $0.3 < \tan^2 \theta_{\odot} < 0.55$ at $2\sigma$~\cite{us:msw2}. 
Several analyzes have arrived to same
conclusions~\cite{balan,fogli,valle2,alia,crem,choubey}.

In a more general context, sub-leading effects can change this
picture, which motivate us to investigate the robustness of the determination 
of the solar parameters.
In this letter, we assume that non-standard neutrino
interactions, which we parameterized by two parameters $\epsilon'$ and
$\epsilon$, are  present, relaxing the allowed region of the
parameters. In the
presence of non-standard neutrino interactions, we have found that the
allowed interval for $\Delta m^2$ increases, rescuing the very low LMA region, 
$\Delta m^2 \sim 1\times 10^{-5}$ eV$^2$, and the high part of LMA region,
$\Delta m^2 \sim 2\times 10^{-4}$ eV$^2$, respectively due the suppression
of earth matter and due to a $\Delta m^2$ shift induced by a non-zero
$\epsilon'$. 

\section{Neutrino Evolution: MSW mechanism and non-standard neutrino
  interactions}

We will work in a two generation neutrino scheme, with the
contribution from the non-standard neutrino interactions
(NSNI)~\cite{nsni0,nsni1,Bergmann:2000gp} added to the usual MSW
Hamiltonian~\cite{MikSmi}. The Hamiltonian in the flavor basis equals 
\begin{eqnarray*} 
H = H_{MSW} + H_{NSNI} & & 
\end{eqnarray*}
where\\
\begin{eqnarray*} 
H_{MSW} &=& 
\left[
\begin{array}{cc}
+\sqrt2 G_F N_e(r) - \frac{\Delta m^2}{4E} \cos{2\theta} &
\frac{\Delta m^2}{4E} \sin{2\theta}  \\
\frac{\Delta m^2}{4E} \sin{2\theta}    &   
\frac{\Delta m^2}{4E} \cos{2\theta} 
\end{array}
\right] 
\end{eqnarray*}
and\\
\begin{eqnarray*} 
H_{NSNI} &=& 
\left[ 
\begin{array}{cc}
0 &
\sqrt2 G_F \epsilon_f N_f(r)  \\
\sqrt2 G_F \epsilon_f N_f(r) &   
\sqrt2 G_F  \epsilon'_f N_f(r)
\end{array}
\right]
\end{eqnarray*}
where $N_f=N_e+2N_n$ when the NSNI occur with d-quarks, 
$N_f=2N_e+N_n$ when u-quarks are involved and simply
$N_f=N_e$ when we have electrons.
The parameters $\epsilon$ and $\epsilon'$ 
describe, respectively, the relative strength of the flavor changing
neutrino interactions and the new flavor diagonal, but
non-universal interactions, normalized to $G_F$.

The NSNI parameters are constrained by non-universal and
flavor-changing processes to be $\epsilon' <0.7$ and $\epsilon
<10^{-2}$~\cite{Bergmann:2000gp}. Since we concentrate on the regions
around LMA neutrino parameters, effects of $\epsilon$ parameters will
be negligibly small and we
can, effectively, set this parameter to zero. We solve numerically the
evolution equation, using the density profile
of the Sun~\cite{ddensity-sun} or the earth~\cite{density-earth}.

We now discuss the behaviour of
$\nu_e$-survival probability that will help to understand our
results. 
In the usual MSW mechanism, for the solar parameters in the LMA region, 
we have a resonant behaviour inside the sun given by 
\be
 \left(\frac{\Delta m^2}{4E} \cos 2\theta\right) \left(\frac{2}{V_0}\right)
 \equiv 
\frac{\Delta m^2 \cos 2\theta}{2\sqrt{2} E G_F N_e} \sim 1~~,
\label{eta}
\ee

In the Sun, $N_n\sim[0.1-0.3]N_e$, and a positive value of 
$\epsilon'$ can be interpreted as a small negative correction in the 
solar density. For a given $\Delta m^2$ and $\theta$, the resonance 
is displaced to the center of the Sun, and as a consequence, less 
neutrinos experience the resonance. As a result, the transition 
between resonant and non-resonant survival probability is displaced
to higher values of $\Delta m^2/4E$, according to:
\[
\frac{\Delta m^2}{4E} \rightarrow \frac{\Delta m^2}{4E} -
\frac{1}{\cos 2\theta}\frac{1}{\sqrt2} G_F {\epsilon'}_{f} N_f(r) ~~.
\]

As stated above, effects of a non-vanishing $\epsilon$
are much weaker, since for LMA we  have $\Delta m^2/(4E) \sim 
10^{-12}$ eV, the same order of $\sqrt2 G_F N_f(r)$ at the resonance
region in the sun. Therefore we should have $\epsilon \sim 1$, much above the
experimental limit, to have some effect on the survival
probability. We neglect $\epsilon$ in what follows.

In the earth $N_e = N_p \sim
N_n \sim {1/3} N_d$. Therefore, for values of $\epsilon'\sim 1/3$ the
matter term in the
evolution matrix due to NSNI has the same order of magnitude of the ordinary 
matter term. As a consequence, the regeneration of $\nu_e$ is
suppressed. 

\begin{figure}[t]
\epsfig{file=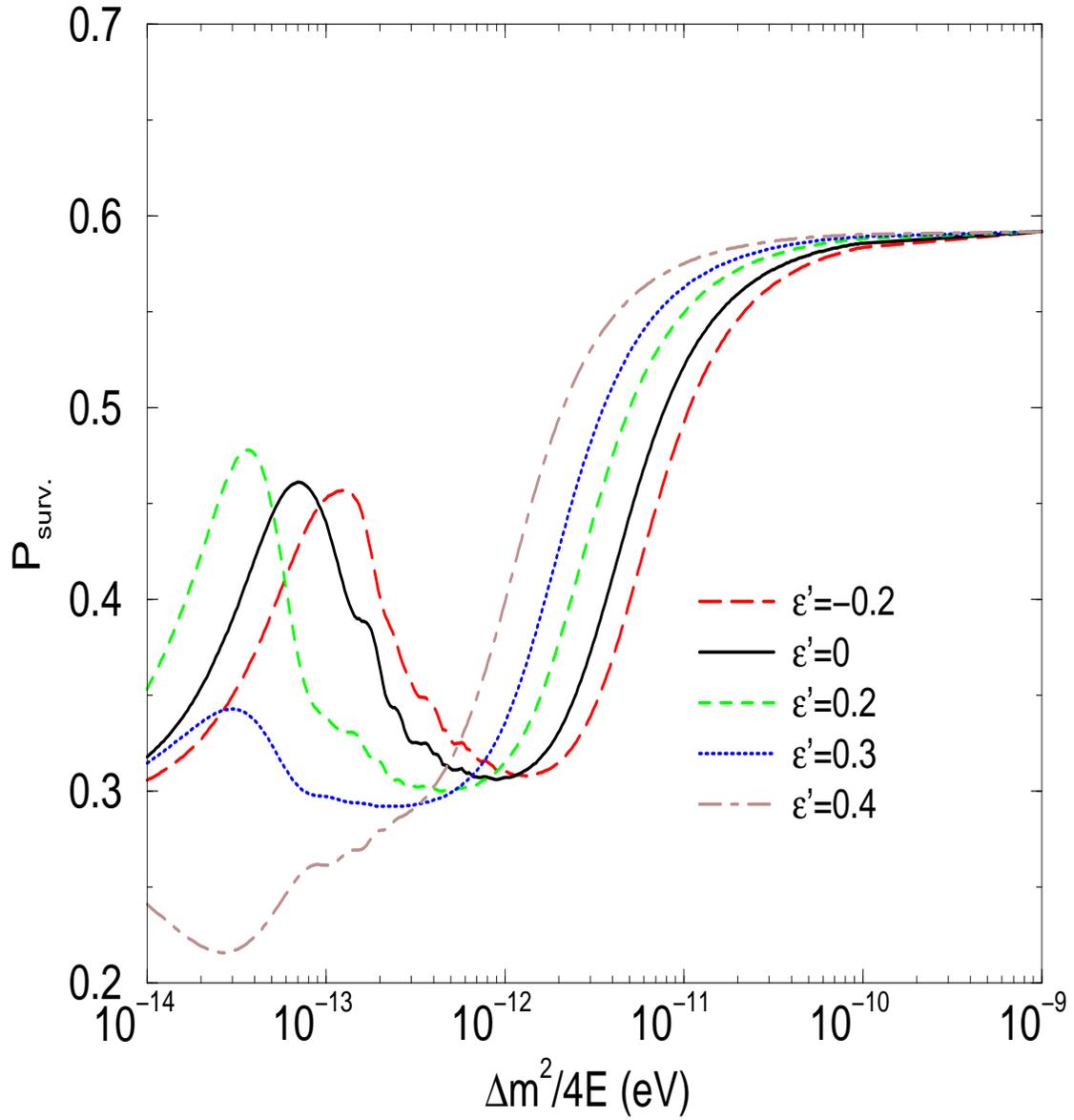, height=16cm,width=15cm}
\caption{Survival probability of electron neutrinos. In this figure we 
can see the displacement in $\Delta m^2/4E$ of the suppression pit 
associated with the transition between resonant and non-resonant regions. 
Also is possible to see the supression of regeneration effect for 
$\epsilon'\sim 0.3$.}
\label{fig:probability}
\end{figure}

Both these effects are presented in Fig.~\ref{fig:probability}. 
Around $\Delta m^ 2/4E \sim [10^{-12}-10^{-11}]$ eV we can see 
the displacement in $\Delta m^2/4E$ of the survival probability in the
Sun, and around 
$\Delta m^ 2/4E \sim [10^{-14}-10^{-12}]$ eV$^2$ the suppression in the
regeneration effect is effective.

%
\section{Solar neutrino and KamLAND data analysis}
We use for the solar neutrino analysis the same data set and the
same procedure of analysis appearing in Ref.~\cite{us:msw2}.  Here the
main ingredients of the analysis are summarized.

The data sample consists of

\noindent

- 3 total rates: (i) the $Ar$-production rate, $Q_{Ar}$, from
Homestake~\cite{Cl}, (ii) the $Ge-$production rate, $Q_{Ge}$ from SAGE
\cite{sage} and (iii) the combined $Ge-$production rate from GALLEX
and GNO \cite{gno};

\noindent

- 44 data points from the zenith-spectra measured by Super-Kamiokande
during 1496 days of operation \cite{SK};

\noindent

- 34 day-night spectral points from SNO plus 
CC, NC and ES rates from SNO salt-phase \cite{sno-all};

\noindent

- 3 fluxes from the SNO salt phase \cite{salt} measured by the CC-, NC
and ES- reactions. 

Altogether the solar neutrino experiments provide us with 84 data
points.
All the solar neutrino fluxes are taken
according to SSM BP2000 \cite{ssm}.

Thus, in our analysis of the solar neutrino data we have three fit
parameters: $\Delta m^2$, $\tan^2\theta_{\odot}$ and $\epsilon'$.\\

We define the contribution of the solar neutrino data to $\chi^2$ as
\begin{equation}
\chi^2_{sun} = \sum_{i,j=1,84}(R^i_{th}-R^i_{ex})\sigma^{-2}_{i,j}
(R^j_{th}-R^j_{ex})~~,
\label{chi-def}
\end{equation}
where we construct the $84\times 84$ covariance matrix $\sigma^2_{i,j}$ 
taking in consideration all correlations between uncertainties.

Following the procedure done in Ref.~\cite{us:kaml}, 
the KamLAND data are analyzed through a Poisson statistics, using 
the following $\chi^2$:

\[
\chi^2_{KL}=\sum_{i=1,13} 2 \left[{N_i^{th}-N_i^{obs} + 
N_i^{obs}ln\left(\frac{N_i^{obs}}{N_i^{th}}\right)}\right]~~,
\label{chisp}
\]
where the $ln$ term is absent when bins with no events are considered (5 last
bins). 

The combined analysis of solar + KamLAND data is done just adding 
the two contributions in $\chi^2$:

\[
\chi^2 = \chi^2_{KL}+\chi^2_{sun}~~.
\]
We minimize the global $\chi^2$ with respect to the three parameters
$\Delta m^2$, $\tan^2\theta_{\odot}$ and $\epsilon'$. For the KamLAND
$\chi^2_{KL}$,
the effect of $\epsilon'$ is negligible due to the short distance
traveled inside earth, then effectively $\chi^2_{KL}$ depends only on 
$\Delta m^2$ and $\tan^2\theta_{\odot}$. We show our results in the plane 
$\Delta m^2$ and $\tan^2 \theta_{\odot}$, in
Fig.~\ref{fig:chi2min}, where we minimized away the
dependence on the $\epsilon'$ parameter. The best fit is for
$\epsilon'=0$, and the usual MSW mechanism is still the best solution
and any non-zero value for $\epsilon'$ parameter only worsens the fit.

To have an idea of the bounds put by the present data and the possible
accumulation of more statistics on KamLAND, we plot in the upper panel
of Fig.~\ref{fig:simula} the
bounds on $\epsilon'$ parameter. The present bound is shown by a thick 
solid curve, where we have $\epsilon'_d<0.3$ at 2$\sigma$. Assuming a
future exposure of 1 kton-yr for KamLAND
experiment and the present solar neutrino data, we simulate the
KamLAND data as generated by a specific $\Delta m^2$ and $\tan^2
\theta_{\odot}$ combination located in each one of three islands: very
low, normal, and high LMA region. If we assume a point in the normal
LMA region, we will get after 1 kton-yr for KamLAND, $-0.4<\epsilon'_d<
0.25$ at $2\sigma$ as showed in the dashed curve. Similar plots for the
high (very low) LMA region as long-dashed (dotted) curves show that the
bounds will be $\epsilon'_d<-0.16$ ($0.16< \epsilon'_d<0.35$). 

\subsection{NSNI with u-quarks and electrons} 

All our results showed  were computed for a NSNI 
with d-quarks. In this section we summarized the
main differences when you have NSNI with u-quarks and electrons.

\begin{figure}[t]
\epsfig{file=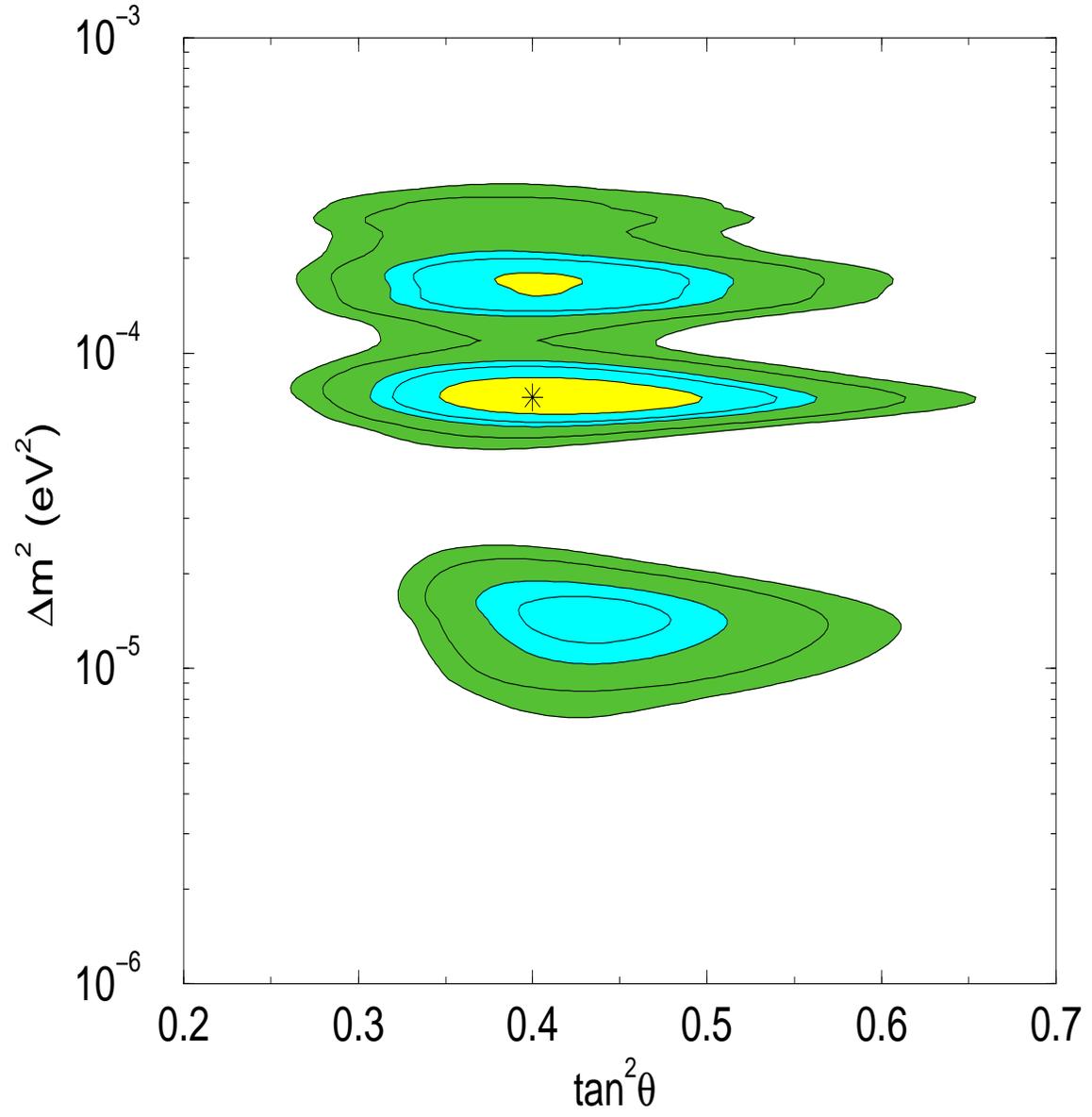, height=16cm,width=15cm}
\caption{Allowed regions of oscillation parameters for 1 $\sigma$, 90\%
  C.L., 95\% C.L., 99\% C. L. and 3 $\sigma$, using the
  constraints of solar+KamLAND data, with NSNI with d-quarks and where
  we minimized the $\chi^2$ with respect to the NSNI parameter
  $\epsilon'$. The best fit point is marked by a star and happens for
  $\epsilon' = 0$.}
\label{fig:chi2min}
\end{figure}

\begin{figure}[t]
\epsfig{file=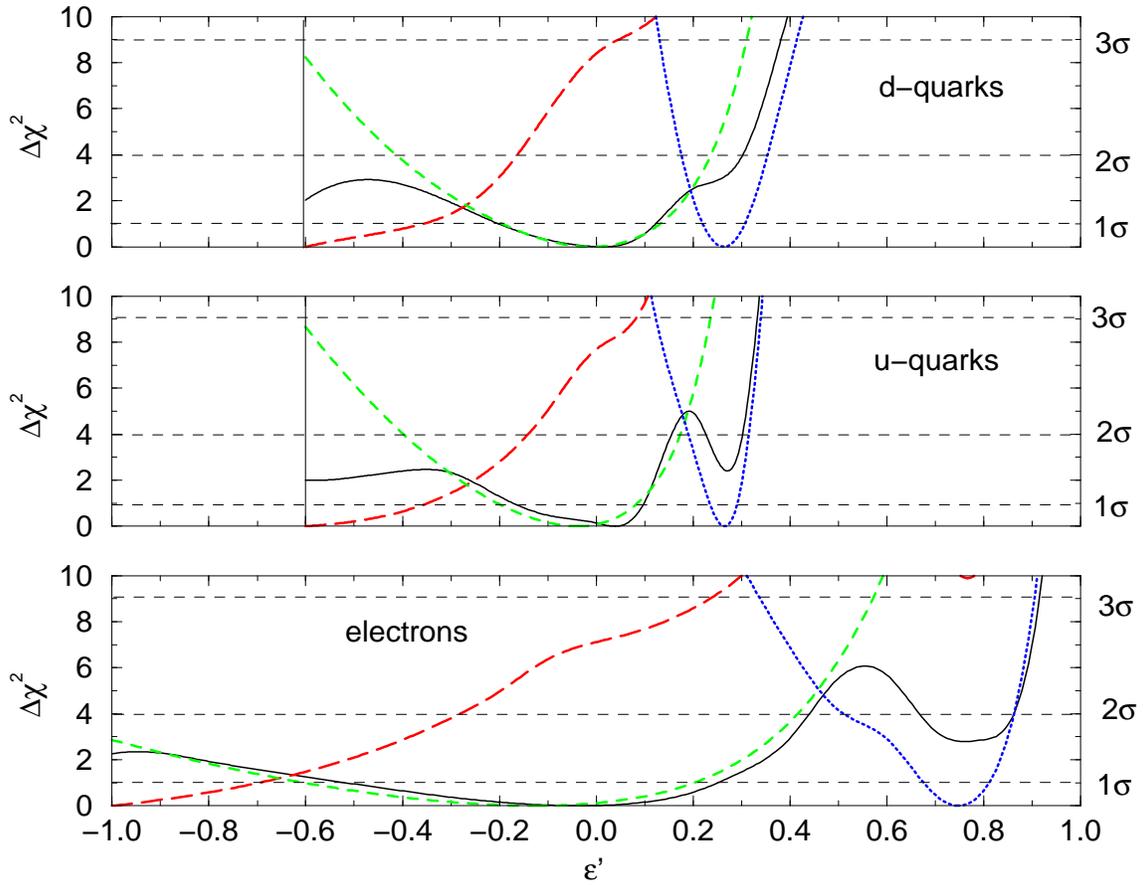, height=12cm,width=15cm}
\caption{Future sensitivity of combined analysis of solar+KamLAND,
  assuming 1kt-yr of exposure. The continuous line corresponds to the
  actual limit that is obtained with present KamLAND data. The dashed,
  dotted and long dashed curves
  refer, respectively, to the simulated parameters $\Delta m^2$ and
  $\tan^2\theta$ lying at low LMA,
  very low LMA, and high LMA regions.}
\label{fig:simula}
\end{figure}

To get NSNI with u-quarks, we need to replace $N_d \rightarrow
N_u$ in the evolution equation for neutrinos. In the production region
where most of $^8B$ neutrinos are created, the ratio between the $N_d$
and $N_u$ densities is almost constant, as can be seen in
Figure 3 of Ref.~\cite{Bergmann:2000gp}. Then the conversion
probability showed in Fig.~\ref{fig:probability} for d-quarks, is similar to
the conversion induced by u-quarks, with the appropriate rescaling of
$\epsilon'_u$ parameter: $\epsilon'_u=\epsilon'_d*N_d/N_u$. 
In the earth the differences are minimal. 
If you compare the allowed region for d-quarks (see
Fig.~\ref{fig:chi2min} ) and u-quarks then we have practically the
same figure for u-quarks.

For NSNI induced by electrons, we can think as a
rescaling of the usual matter potential of the MSW mechanism: 
$\sqrt2 G_F N_e(r) \rightarrow  \sqrt2 G_F N_e(r)- \sqrt2 G_F
{\epsilon'}_{e} N_e(r)= \sqrt2 (1-{\epsilon'}_{e}) G_F
N_e(r)$. 
The parameter $\epsilon'_e$ have looser bounds then
$\epsilon'_d$~\cite{Guzzo:2000kx} and values for $\epsilon'_e \sim 1$
are still allowed. 
One could worry that such large values of $\epsilon'_e$ could cause a
strong effect on the detection cross section $\sigma (\nu e
\rightarrow \nu e) $ used to detect solar neutrinos on  Super
Kamiokande experiment.
This is not true because the matter potential 
induced by $\epsilon'_e$ is proportional only to the 
vector contribution of the non-standard neutrino couplings, 
and the cross section depends on a combination of the left/right couplings of
the non-standard neutrino interaction of the electron neutrino.

When we looked for the allowed regions, we have similar figures as in
Fig.~\ref{fig:chi2min} and we obtain again the appearance of a very
low LMA region. We have also shown in Fig.~\ref{fig:simula} 
(second and third panels) the limits in $\epsilon'$ for NSNI
induced by u-quarks and electrons. For comparison, similar plot was
obtained in Ref.~\cite{Fogli:2003vj}, bu. 

\section{Discussion of results}

In the allowed region showed, we notice two distinguished facts: the
appearance of new very low LMA region at $\Delta m^2 \sim 10^{-5}$
eV$^2$ and
the old very high LMA, $\Delta m^2 \sim 1\times 10^{-4}$ eV$^2$, appears at
$ 1 \sigma$. 

The analysis of the data, in the absence of $\epsilon'$ parameter,
disfavor the high LMA solution that is only allowed at
$3\sigma$~\cite{us:msw2}. 
Assuming a negative $\epsilon'$, the situation changes due the shift
of the resonance layer in the sun, as commented before. For 
$\epsilon'=-0.6$ (the lower limit we used in our analysis) this region
is  now accepted at $1\sigma$ C.L.. 

The most interesting phenomenology happens at the island of 
allowed region in KamLAND analysis around 
$\Delta m^2\sim 10^{-5}$~eV$^2$. This region is not allowed 
in a pure-LMA scenario due to the very high day-night asymmetry 
that is expected for these parameters. But when we include a
$\epsilon'>0$, the NSNI term in Hamiltonian compensates the effect of
Earth matter interaction. For 
$\epsilon'\sim 1/3$ we expect a very low regeneration, and for 
larger values of $\epsilon'$ we can have even a positive day-night 
asymmetry. 
This region is also allowed when we assume other non-standard
mechanisms, as random matter density fluctuations in the sun, as
presented in~\cite{Guzzo:2003xk}.

\section{Conclusions}

We showed that NSNI will affect the fit in
the LMA region of the MSW solution to the solar neutrino anomaly. When
one takes into account the KamLAND results, positive values of the
$\epsilon'$  push the allowed region of the neutrino parameters
$\Delta m^2$ and $\tan^2\theta_{\odot}$ at 95\% C.L. from pure MSW low-LMA
and high-LMA to a completely new region in which $\Delta m^2$ is
lower than the previous two ones, which we call very-low-LMA. If one chooses
$\epsilon'<0$, the preferred allowed region tends to higher values of
$\Delta m^2$.  

Almost all our conclusions below are independent of specific sources
of the non-standard neutrino interactions, that could be present in
interactions with d-quarks, u-quarks or electrons. 

We have found that the main effects of the presence of the NSNI
interactions are:

\begin{itemize}

\item{displacement of low-LMA region to lower (higher) values of $\Delta m^2$, 
for a positive (negative) value of $\epsilon'$.}

\item{suppression of Earth regeneration at $\Delta m^2\sim 10^{-5}$ 
eV$^2$ for positive values of $\epsilon'$.}

\item{Due to suppression of Earth regeneration, appearance of a new region 
of compatibility between solar and KamLAND data around $\Delta m^2\sim  
10^{-5}$ eV$^2$,  with no spectrum distortion for the low-energy
  SK and SNO data.}

\item{improvement of high-LMA fit quality for positive values of $\epsilon'$}

\item a 1 kton-yr of KamLAND can make a strong statement about the
existence of non-standard neutrino interactions. The striking signal
of this NSNI would be the location of the prefered oscillation
parameters in the very low or in the high LMA region. 
\end{itemize}

{\em Note added:} When we were finishing our paper, an 
article by Friedland, Lunardini and Pe\~na-Garay (hep-ph/0402266)
appeared, 
which discusses topics similar to the ones discussed in our paper,
where we discuss not only the non-standard neutrino interaction induced by
d-quarks case as well the u-quarks and electrons. Also we made a
quantitative statement about the role of more statistics on KamLAND
experiment, combined with the present solar neutrino data, 
to put more restrictive bounds on non-standard neutrino
interactions. 

\begin{acknowledgments} 
This work was supported by Funda\c{c}\~ao de Amparo
 \`a Pesquisa do Estado de S\~ao Paulo (FAPESP) and  Conselho
Nacional de Desenvolvimento Cient\'\i fico e Tecnol\'ogico (CNPq).
\end{acknowledgments}

\end{document}